# Application of artificial intelligence techniques for automated detection of myocardial infarction: A review

Javad Hassannataj Joloudari[1], Sanaz Mojrian[2], Issa Nodehi[3], Amir Mashmool[4], Zeynab Kiani Zadegan[1], Sahar Khanjani Shirkharkolaie[2], Roohallah Alizadehsani[5], Tahereh Tamadon[1], Samiyeh Khosravi[1], Mitra Akbari Kohnehshari[6], Edris Hassannatajjeloudari[7], Danial Sharifrazi[8], Amir Mosavi[9], Hui Wen Loh[10], Ru-San Tan[11], U Rajendra Acharya[10,12,13]

[1]Department of Computer Engineering, Faculty of Engineering, University of Birjand, Birjand, Iran
[2]Department of Information Technology Engineering, Mazandaran University of Science and Technology, Babol, Iran
[3]Department of Computer Engineering, University of Qom, Qom, Iran
[4]Department of Informatics, Bioengineering, Robotics and System Engineering (DiBRIS), University of Genova(UniGE), Italy
[5]Institute for Intelligent Systems Research and Innovation, Deakin University, Geelong, VIC 3216, Australia
[6]Computer Engineering Department, Engineering Faculty, Bu-Ali Sina University, Hamedan, Iran
[7]Department of Nursing, School of Nursing and Allied Medical Sciences, Maragheh Faculty of Medical Sciences, Maragheh, Iran
[8]Department of Computer Engineering, Shiraz Branch, Islamic Azad University, Shiraz, Iran
[9]Faculty of Informatics, Obuda University, Budapest, Hungary
[10]School of Science and Technology, Singapore University of Social Sciences, Singapore
[11]Department of Cardiology, National Heart Centre Singapore, Singapore
[12]Ngee Ann Polytechnic, Singapore, 599489, Singapore
[13]Dept. of Biomedical Informatics and Medical Engineering, Asia University, Taichung, Taiwan

*Abstract*— Myocardial infarction (MI) results in heart muscle injury due to receiving insufficient blood flow. MI is the most common cause of mortality in middle-aged and elderly individuals around the world. To diagnose MI, clinicians need to interpret electrocardiography (ECG) signals, which requires expertise and is subject to observer bias. Artificial intelligence-based methods can be utilized to screen for or diagnose MI automatically using ECG signals. In this work, we conducted a comprehensive assessment of artificial intelligence-based approaches for MI detection based on ECG as well as other biophysical signals, including machine learning (ML) and deep learning (DL) models. The performance of traditional ML methods relies on handcrafted features and manual selection of ECG signals, whereas DL models can automate these tasks. The review observed that deep convolutional neural networks (DCNNs) yielded excellent classification performance for MI diagnosis, which explains why they have become prevalent in recent years. To our knowledge, this is the first comprehensive survey of artificial intelligence techniques employed for MI diagnosis using ECG and other biophysical signals.

## I. INTRODUCTION

In myocardial infarction (MI), or heart attack, heart muscle cells die from lack of oxygen due to insufficient blood supply [1-5]. The latter is predominantly caused by coronary artery disease, in which the lumina of coronary arteries supplying the heart muscle become stenotic from atherosclerosis of the artery walls. In advanced coronary artery disease, the atherosclerotic plaque expands and becomes vulnerable to surface rupture [6], which can trigger the sudden formation of lumen-occluding thrombus, resulting in MI. This typical MI scenario is depicted in Fig.1, where the death of a region of the heart muscle is caused by acute thrombus occlusion adjacent to a ruptured cholesterol-laden plaque at the site of coronary artery stenosis (inset).

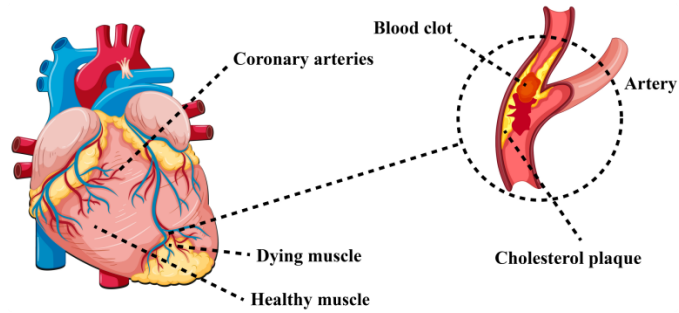

Fig. 1. Illustration of myocardial infarct caused by acute thrombus in the culprit coronary artery (see text).

After a MI, the left ventricle enlarges and undergoes functional changes in response to injury. This eventually leads to congestive heart failure, where the weakened heart muscle is unable to pump blood through the body efficiently, and at the advanced stage, death can ensue. Prompt diagnosis of MI and early intervention are critical for patients' survival. Electrocardiography (ECG) is the most common method used to diagnose [7-9]. ECG is a surface map of the underlying cardiac action potentials during electrical signal conduction through the heart chambers. It will typically show abnormal morphology at lead positions overlying the involved MI region. The open-access Physikalisch-Technische Bundesanstalt (PTB) database is a repository of cardiologist-annotated ECGs of diverse cardiological conditions, including MI, which scientists frequently use for MI research.

Other methods for diagnosing MI include noninvasive imaging, e.g., echocardiography and magnetic resonance imaging, as well as clinical and serological parameters. The manual interpretation of ECG, imaging readouts, and clinical parameters require expertise and may be subject to intra- and inter-observer variability. Artificial intelligence (AI)-enabled automated computer-aided diagnostic systems [10-13] may improve efficiency and reduce observer bias of screening for MI using the different biological signals.

Both machine learning (ML) and deep learning (DL) models may be deployed for discriminating MI vs. normal at the signal readout or subject levels. In ML [14], feature extraction and classification are separate sequential operations that may require high-level handcrafted engineering decisions. In contrast, in DL [15], feature extraction and classification are integrated and automatically performed [16] (Fig. 2). A DL model typically comprises some form of artificial neural network (ANN) with many hidden layers that can automatically extract prominent features from high-dimensional raw data (e.g., images) [17-21]. An example of the DL model is the deep convolutional neural network (DCNN) [20, 22], which may have ten to hundreds of hidden layers [23], including several convolutional, pooling, and fully-connected layers (Fig. 3). The input signals are convolved by the convolutional kernels to extract features. The pooling layer reduces the network's computational complexity while maintaining a consistent feature map resolution. Two well-known types of pooling layers are max-pooling and average pooling. The last layer of the DCNN is a fully-connected layer that outputs the final classification results. DL models usually yield excellent performance for detecting and classifying early changes in the disease course [23-31].

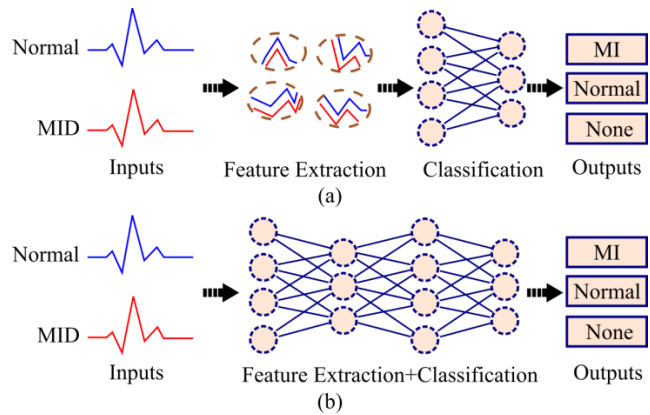

Fig. 2. Illustration of machine learning and deep learning architectures for automated detection of MI.

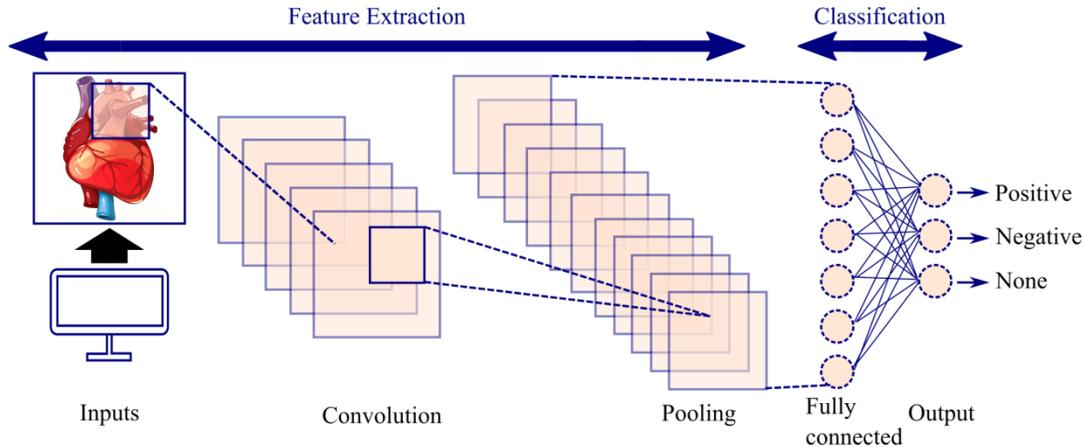

Fig. 3. Conventional deep convolutional neural network structure.

This paper has performed a systematic review of recent studies on artificial intelligence for MI diagnosis, stratified by ML- or DL-based approaches. The rest of the paper is structured as follows: search methodology is presented in Section 2; literature review on AI for MI diagnosis, in Section 3; results and discussion, in Section 4; future works, in Section 5; and conclusion, in Section 6.

## II. SEARCH STRATEGY

We performed a search for works published between January 1st, 1992 to January 31st, 2022, on the Google Scholar engine using the keywords "myocardial infarct diagnosis", "artificial intelligence", "machine learning" and "deep learning". As a result, we retrieved 66 papers (31 and 35 related to ML and DL, respectively), most of which were published by IEEE, Elsevier, and Springer (Fig. 4).

## III. A SYSTEMATIC LITERATURE REVIEW FOR MI DIAGNOSIS

### A. ML-based methods

ML approaches for MI classification include k-nearest neighbor (KNN), decision tree (DT), support vector machine (SVM), naïve Bayes (NB), and random forest (RF) as well as ANN which is inspired by human neuronal function [32]. Readdy et al. [33] used 15 features of the V2 to V4 chest lead ECG QRS measurements in their ANN-feedforward MI classification model and attained 79% accuracy and 97% specificity. Hedén et al. [34] applied ANN classification to 1,120 and 10,452 MI and normal ECGs, respectively, and achieved 95.0% sensitivity and 86.3% specificity. Lu et al. [35] combined fuzzy logic theory and backpropagation neural network (BPNN) to create a neuro-fuzzy FL-BPNN classification model that attained 89.4% and 95.0% accuracy rates for MI and normal subjects, respectively. Haraldsson et al. [36] used Bayesian ANN trained with Hermite expansion coefficients BANN-HE to construct a 12-lead ECG-based MI diagnostic system that showed strong discriminative utility for MI vs. normal (C statistic 83.4%) on 2,238 ECG signals of emergency department attendance.

Zheng et al. [37] studied comprehensive192-lead body surface potential maps and used SVM, NB, and RF classifiers to diagnose MI, achieving accuracy rates of 82.8%, 81.9%, and 84.5%, respectively. Arif et al. [38] proposed a BPNN method using principal component analysis (PCA) to extract features, which achieved 93.7% classification accuracy for MI diagnosis. Sun et al. [1] described a method for diagnosing MI using 12-lead ECGs called latent topic multiple instance learning (LTMIL). Signal processing was done with discrete cosine transform (DCT) bandpass filters. Fifth-order polynomial fitting was utilized to establish the 74-dimensional feature spaces. A particle swarm optimizer was used for variable weighting. SVM, KNN, RF, and ensemble learning were utilized for classification. KNN ensemble combined with LTMIL achieved the highest accuracy of 90%. Arif et al. [39] diagnosed MI using the KNN method on 20,160 ECG beats obtained from the PTB database. The experimental phase used 10,080 and 711 heartbeats for non-pruning and pruning training, respectively. Dual wavelet transform was applied to the ECG signals to determine the 36 components of the feature vector. Finally, MI cases were divided into 11 classes (10 classes for the various infarct sites and one class for normal subjects). They attained 98.8% overall classification accuracy and sensitivity and specificity exceeding 90%.

Chang et al. [40] used four chest ECG leads (Leads V1, V2, V3, and V4) to diagnose MI with hidden Markov model (HMM), Gaussian mixture model (GMM), SVM, and Viterbi algorithm. On a 582 MI and 547 normal heartbeats dataset, the combination of HMM and GMM achieved the best accuracy of 82.50% for MI diagnosis. For the detection and localization of MI, Safdarian et al. [41] studied classification approaches such as probabilistic neural network (PNN), KNN, multilayer perceptron (MLP), and NB. They used the NB classifier to obtain 94.74% accuracy for MI detection and the PNN approach to achieve 76.67% accuracy for MI localization.

Kora et al. [42] used an improved bat algorithm (IBA) to extract the major properties of each pulse from the PTB database, which included 148 MI and 52 normal individuals. Backpropagation Levenberg–Marquardt Neural Network (LMNN) classifier was used to input the best features. The combination of optimized features, IBA and LMNN achieved 98.9% accuracy for MI diagnosis, outperforming methods like SVM, scalar conjugate gradient neural network, LMNN, and KNN. For MI diagnosis, Sharma et al.

[43] used a multiscale energy and eigenspace technique. After applying wavelet decomposition of multi-lead ECG signals to clinical components in various subgroups, a frame with four beats from each ECG lead was utilized to detect MI. Multilayer ECG frames were used to adjust the properties of the 72-dimensional vectors of 12-lead ECG data. The ECG signals were classified using SVM with radial basis function (RBF) kernel, linear SVM, and KNN, which attained 96.0% accuracy for MI diagnosis.

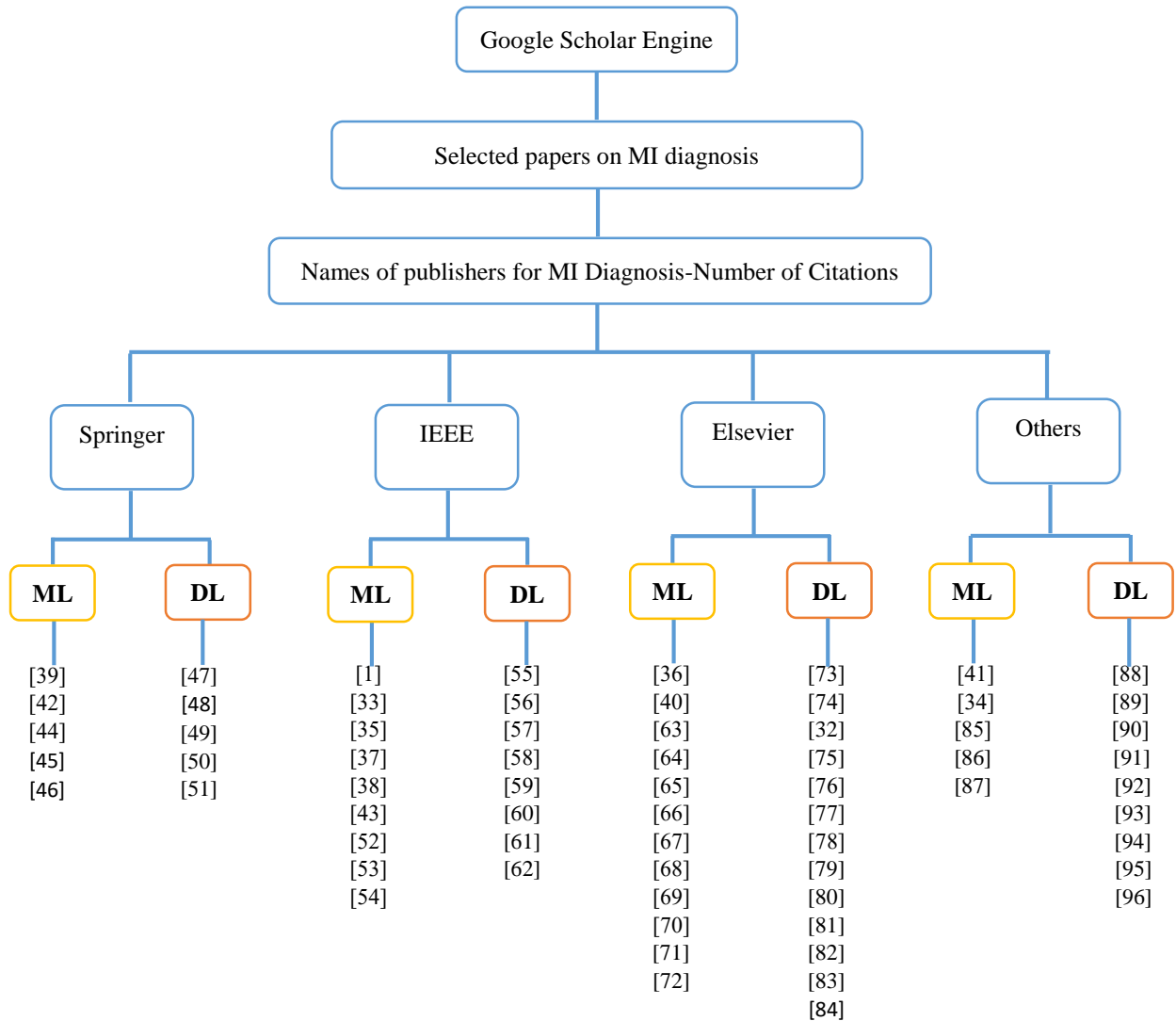

Fig. 4. Systematic literature review of the machine and deep learning methods for myocardial infarct diagnosis. The numbered relevant references are in square parentheses. DL, deep learning; ML, machine learning.

In the PTB database, Acharya et al. [63] used a KNN classifier to differentiate MI vs. normal ECGs in the PTB database. Each signal underwent four levels of discrete wavelet transform (DWT) decomposition using Daubechies' six wavelet basis function, and 12 types of nonlinear properties were extracted from the DWT coefficients. The discriminative features ranked based on their t-values, F-values, and analysis of variance (ANOVA) were used to derive the rankings of the normal class and ten types of MI. The method achieved 98.80% classification accuracy for MI vs. normal classes based on 47 characteristics obtained from Lead V5. Moreover, 98.74% accuracy for 11-class classification based on 25 characteristics from Lead V3 and 99.97% accuracy for MI localization based on Lead V3 was achieved. In another study, Acharya et al. [64] compared DWT, empirical mode decomposition (EMD), and DCT methods for coronary artery disease and MI diagnoses. ECG signals underwent DCT, DWT, and EMD to obtain the corresponding coefficients, which were then reduced using the locality preserving projection (LPP) method. The LPP features were ranked using their F-values, highly ranked coefficients were then fed to the KNN classifier. DCT coefficients paired with KNN yielded the maximum accuracy of 98.5%.

To diagnose MI using ECG data, Kumar et al. [85] used a sample entropy in a flexible analytical wavelet transform (FAWT) framework. FAWT was utilized to break down each ECG beat into sub-band signals after the ECG signals were split into beats. These sub-band signals were used to calculate sample entropies, which were then input into several classifiers. The classification accuracy of FAWT combined with least-squares SVM (LS-SVM) was 99.31%, outperforming RF, J48 decision tree, and BPNN techniques. Khatun and Morshed [52] developed a bagging trees (BTs) classification method that could automatically identify P,

Q, R, S, and T waves on single-lead ECGs and extract 33 features of 15 interval and 18 amplitude types. The method was able to diagnose MI with 99.7% accuracy using ECG Lead V4. Acharya et al. [65] proposed a KNN classification model to diagnose MI using Lead II ECG signals acquired from the PTB database. The modified binary particle swarm optimization method was used to identify informative aspects of ECG signals, which were then ranked using ANOVA and relief methods. The top-ranked features were then fed to the DT and KNN classifiers. 99.55% and 99.01% accuracy rates were obtained using the KNN method combined with contourlet and shearlet transform, respectively, based on 20 selected features of the ECG signals.

Dohare et al. [66] employed a 12-lead ECG signal and combined SVM with a PCA reduction methodology to identify MI. Each ECG lead was examined with the help of the composite lead, and the PCA method was used to minimize computing complexity and the number of features. 98.33% and 96.66% accuracy rates were attained on the original feature set vs. PCA-extracted features, respectively. Diker et al. [53] combined SVM with a genetic algorithm (GA) to diagnose MI on ECGs from the PTB database. 23 morphological, time-domain, and DWT features were extracted from ECG signals, and their dimensionality was reduced to 9 using GA. The SVM classifier attained 87.8% and 86.44% accuracy rates using the reduced 9 and original 23 features, respectively. Han and Shi [67] explored methods such as SVM (with RBF, linear, and polynomial kernels), BTs, and BPNN for MI diagnosis and developed a hybrid feature set for ECG signals composed of energy entropy as global features and local morphological features. The global features were computed using maximal overlap discrete wavelet packet transform (MODWP) of ECG signals. After the fusion of multi-lead ECG signals, PCA, linear discriminant analysis, and LPP approaches were employed to reduce the number of features. SVM-RBF with 10-fold cross-validation (CV) achieved the greatest accuracy of 99.81 percent utilizing the 18 features for the intra-patient pattern in trials using ECGs from the PTB database.

Zhang et al. [54] proposed staked sparse autoencoder (SAE) with a tree bagger (TB) for diagnosing MI using single-lead ECG signals of the PTB database. To avoid the vanishing gradient problem in the feature extraction network, SAE employs a layer-wise training technique. Without an input tag, it may learn the best feature expression from the heartbeat. As a result, unique characteristics can be extracted from single-lead ECG signals using this method. By merging the findings of numerous decision trees and feature improvements, the TB classifier was created to best simulate MI diagnosis. The model attained accuracy, sensitivity, and specificity of 99.90%, 99.98%, and 99.52%, respectively. Zeng et al. [68] used 12-lead and Frank XYZ lead ECG signal segments from the PTB database to propose a neural network with RBF for early MI diagnosis. To develop cardiac vectors based on the synthesis of 12-lead ECG signals and Frank XYZ leads, nonlinear feature extraction methods such as tunable quality factor wavelet transformations, variational mode decomposition, and phase space reconstruction were applied. To model, diagnose, and classify MI vs. healthy people, these feature vectors were fed into dynamical estimators incorporating an RBF-neural network. The method attained the best accuracy of 97.98% using a 10-fold CV. Kayikcioglu et al. [69] deployed ensemble classification algorithms such as boosted trees, BTs, and subspace KNN in addition to standard SVM and KNN algorithms for MI diagnosis using ECGs from the MIT-BIH Arrhythmia, European ST-T, and the Long-Term ST databases. The datasets were classified into four classes: healthy, arrhythmia, ST depression, and ST elevation (ST changes can be present in MI). Quadratic time-frequency distributions including smoothed Wigner-Ville, the Choi-Williams, the Bessel and the Born-Jordan were applied on 5-lead ECG signals for feature extraction. The best accuracy of 94.23% was obtained for the weighted KNN algorithm using features extracted by the Choi-Williams time distribution.

Liu et al. [70] proposed a single-beat MI diagnosis system based on the ECG denoising algorithm dual-Q tunable Q-factor wavelet transformation (Dual-Q TQWT). The proposed Dual-Q TQWT and wavelet packet tensor decomposition (WPTD) were superior to DWT and EMD denoising methods in their experiments. Using the Dual-Q TQWT + WPTD + multilinear PCA + TB system, they achieved 99.98% accuracy in beat level and 97.46% accuracy in record level for classification of MI vs. normal. Lin et al. [44] decomposed ECG signals using MODWP and extracted features such as variance, interquartile range, Pearson correlation coefficient, Hoeffding's D correlation coefficient, and Shannon entropy of the wavelet coefficients. Inputting these features to a KNN classifier, they attained 99.57% accuracy for MI diagnosis using ECGs in the PTB database.

Valizadeh et al. [45] proposed a novel parametric-based feature selection based on the left ventricle's 3D spherical harmonic shape descriptors to distinguish MI patients from healthy ones. The method was based on the hypothesis that spherical harmonic coefficients of the parameterized endocardial shapes would be different for MI patients and healthy subjects. The method was started by preprocessing cine-MRI images from the automated cardiac diagnosis challenge dataset. Next, parametric-based features, i.e., spherical harmonic coefficients, were extracted and normalized. PCA was applied to the normalized features, and the results were used to train multiple classifiers, among which SVM attained the best performance.

### B. DL-based methods

DL can learn huge datasets due to their complex structure with several layers. DL models such as DCNN, long short-term memory (LSTM), recurrent neural network (RNN), and autoencoder network can be used for disease classification and generally outperform ML methods in terms of signal processing and classification [97]. Acharya et al. [73] applied an 11-layer DCNN on 10,546 and 40,182 Lead II ECG signals of normal and MI, respectively, and reported 93.53% and 95.22% accuracy rates on noisy and noiseless data, respectively. Reasat and Shahnaz [55] designed a DCNN to identify inferior MI ECG signals from healthy ones using ECG Leads II, III, and AFV inputs. They tested their network on data from one patient and trained it on data from the rest of the patients. The model attained 84.54% accuracy, outperforming stationary wavelet transform (SWT) with KNN and SWT with SVM [98]. Lui and Chow [76] used a DCNN combined with LSTM stacking decoding RNN to diagnose MI using Lead I ECG signals. They attained 92.4% sensitivity, 97.7% specificity, 97.2% positive predictive value, and 94.6% F1 score, which were superior to the performance of pure DCNN with MLP.

Gupta et al. [47] observed that data from ECG Leads V6, V3, and II were critical for identifying MI correctly and applied this insight to modify the ConvNetQuake neural network for MI classification. The modified model achieved 99.43% accuracy for MI diagnosis using only 10 seconds of raw ECG recordings as input. Baloglu et al. [77] presented an end-to-end DCNN model for MI diagnosis using 12-lead ECG signals. The model attained 99.78% accuracy on ECG Lead V4. Tripathy et al. [56] presented a novel MI diagnostic approach that combined deep layer LS-SVM with features obtained by time-scale decomposition of 12-lead ECG signals using Fourier–Bessel series expansion-based empirical wavelet transform. The system achieved 99.74% accuracy. Zhang et al. [57] used Gramian angular difference field (GADF), PCA network (a lightweight DCNN-like model), and linear SVM in combination to extract salient features of Lead II ECGs from the PTB database. For ECG-level classification, the model achieved 98.44% (beat type: no noise) accuracy rates with 5-fold CV. At the subject-level, 93.17% accuracy was achieved.

Feng et al. [88] proposed a multichannel classification algorithm that combined a 16-layer DCNN with LSTM for MI diagnosis. ECG signals were preprocessed to extract heartbeat segments, and the extracted segments were then fed to the DCNN to obtain the feature map. Final classification results were output by LSTM based on the received feature map. The system attained 95.4% accuracy without the use of handcrafted features. Liu et al. [58] combined DCNN and RNN to build a hybrid network named multiple-feature-branch convolutional bidirectional RNN (MFB-CBRNN) for MI diagnosis using 12-lead ECG signals. The RNN was used to summarize the characteristics of the 12-lead ECG records. The model attained 99.90% and 93.08% accuracy rates at the ECG and subject levels, respectively. Strodthoff and Strodthoff [89] used an ensemble of fully-convolutional DCNNs for MI diagnosis using 12-lead ECGs from the PTB database and reported 93.3% sensitivity and 89.7% specificity using 10-fold CV.

Han and Shi [74] used 12-lead ECG signals from the PTB database to create a multi-lead residual neural network (ML-ResNet) model with three residual blocks and feature fusion for MI diagnosis. The model attained 95.49% and 99.92% accuracy rates for the inter-and intra-patient schemes, respectively. Natesan et al. [59] used multi-lead ECG signals from the PTB database to classify MI using DCNN with data augmentation, without data augmentation, and with noise, achieving 94.98%, 90.34%, and 90.93% accuracy rates, respectively. Fu et al. [90] developed a multi-lead attention mechanism (MLA-CNN-BiGRU) framework for diagnosing MI using 12-lead ECG signals from the PTB database. The model performance was enhanced by weighting the different leads in proportion to their contributions. In addition, interrelated characteristics between leads were exploited to extract discriminative spatial features using the two-dimensional DCNN module. With the memory capability of BiGRU, the model was able to exploit the temporal features of ECG signals, and a combination of temporal and spatial features was used for classification. The model achieved 99.93% and 96.5% accuracy rates for intra- and inter-patient schemes, respectively.

Tadesse et al. [75] presented an end-to-end DL approach to diagnose MI and categorize its time of occurrence as acute, recent, or old. Being able to delineate the time of MI occurrence has implications for the timelines of therapeutic intervention, especially in acute cases. The model's computational complexity was reduced by utilizing transfer learning based on existing pre-trained networks. As a result, the model attained fair to good discriminative performance with C-statistics of 96.7%, 82.9%, 68.6%, and 73.8% reported for the normal, acute, recent, and old MI classes, respectively. Jahmunah et al. [32] compared the performance of DCNN vs. Gabor-filter DCNN models for classifying subjects into MI, coronary artery disease, congestive heart failure, and normal classes. In the latter, eight convolutional layers of the DCNN were replaced with Gabor filters, which reduced the computational complexity. Based on Lead II ECG signals, the Gabor-filter DCNN and DCNN models attained average accuracy rates of 99.55% and 98.74%, respectively, for the four-class classification task.

Kim et al. [78] utilized U-Net architecture combined with the dropout technique to estimate the uncertainty of the U-Net model using cardiac perfusion images for myocardial segmentation. Their approach obtained better Dice similarity of 0.806± 0.096 (average ± standard deviation) compared to rival methods such as semi-automatic U-Net (0.808±0.084) and automatic U-Net (0.729±0.147).

Garland et al [91] studied the possibility of using CNNs to distinguish MI subjects from healthy ones. To this end, the classification performance of four different CNNs (commonly used in surgical/anatomical histopathology) was investigated on a dataset with 150 images (50 normal myocardium, 50 acute MI, and 50 old MI). The authors reported that InceptionResNet v2 with 95% accuracy was a promising candidate for MI diagnosis. As another MI diagnosis study based on non-ECG data, Degerli et al [61] gathered an echocardiographic dataset (HMC-QU) for MI detection, which was publicly available. They proposed a three-phase approach to early MI detection. The first phase involved using DL to segment the left ventricle. Next, the segmented region was analyzed for feature engineering. Finally, in the third phase, MI detection was performed.

As mentioned before, precise and timely MI identification is critical for patients' survival. Myocardial contrast echocardiography (MCE) has been used in MI diagnosis but is time-consuming, subjective, and highly operator-dependent. In [79], a new DL network named polar residual network (PResNet) based on ResNet was proposed for automated computer-aided MI diagnosis based on MCE images. The authors defined a new polar layer in the structure of PResNet that mapped subsections of MCE to the polar map. The rest of the convolutional and residual layers of the networks were used to extract salient features from the polar layer.

*C. Public ECG datasets for MI*

ECG is a key non-invasive approach for cardiovascular diseases diagnosis and the research community can benefit from high-quality and publicly available ECG datasets. One famous ECG datasets is the Physikalisch-Technische Bundesanstalt (PTB) diagnostic ECG dataset [73, 99]. Publicly available for over 20 years, it has been used in various studies on MI diagnosis, including several in this review. More recently, Wagner et al. [100] released one of the largest ECG datasets named PTB-XL. Access to this dataset had previously been limited but was recently for public use in 2020. The dataset comprises 21,837 10-second 12-lead ECG records from 18,885 patients (52% male, 48% female; median age 62 years, range 0 to 95 years) with diverse diagnoses. In the PTB-XL, 12-lead ECG of 148 MI patients and 52 healthy subjects can be used for training MI diagnosis models. Another publicly available ECG dataset is the MIT-BIH Arrhythmia database, which has been widely used for the classification of cardiac arrhythmia. A short description is given here as it has been used for arrhythmia classification in experiments conducted by some of the reviewed papers. MIT-BIH consists of 48 half-hour excerpts of two-channel ambulatory ECG recordings of 47 subjects acquired between 1975 and 1979 from inpatients (about 60%) and outpatients (about 40%) at the Beth Israel Hospital, Boston. Twenty-three recordings were randomly chosen from 4000 24-hour ambulatory ECG recordings as well as 25 other recordings from patients with less common but clinically significant arrhythmia selected from the same set [63, 101].

## IV. RESULTS AND DISCUSSION

The results of the ML-based and DL-based methods are summarized in Tables 1 and 2. From Table 1, among ML models, SAE+TB proposed by Zhang et al. [54] attained the best accuracy of 99.90% using the PTB database. From Table 2, the DCNN method has the highest accuracy of 99.95% for MI diagnosis using the PTB database.

**TABLE 1.**
Summary of machine learning-based publications on myocardial infarct diagnosis

| No. | References | Methods | No. K-fold CV | Dataset | ACC (%) |
|---|---|---|---|---|---|
| 1 | Readdy et al., [33] | ANN | NC | Leads: V2-V4 Subjects: 272 MI, 479 Normal | 79 |
| 2 | Hedén et al., [34] | ANN | 8-fold CV | Leads: 12 leads Subjects: 1120 MI, 10452 Normal from PTB database | N/A |
| 3 | Lu et al., [35] | FL-BPNN | NC | Leads: 12 leads, subjects: 20 normal, 104 MI | 89.40 |
| 4 | Haraldsson et al., [36] | ANN | 3-fold CV | Leads: 12 leads subjects: 2238 ECGs, 699 men and 420 women for MI group, 578 men and 541 women for Normal group | 94 |
| 5 | Zheng et al., [37] | Random Forest | 10-fold CV | Leads: 192-lead body surface potential maps Subjects: 116; 57 MI, 59 Normal from PTB database | 84.50 |
| 6 | Arif et al., [38] | BPNN+ PCA | NC | Leads:12 leads Subject:148 MI and 52 Normal from PTB database | 93.70 |
| 7 | Sun et al., [1] | KNN ensemble+LTMIL | 10-fold CV | Leads:12 leads Subject: 369 MI, 79 Normal from PTB database | 90 |
| 8 | Arif et al., [39] | KNN | 10-fold CV | Leads: 12 leads Subjects: 10 types of MI from PTB database | 98.30 |
| 9 | Chang et al., [40] | HMMs + GMMs | NC | Leads: Leads V1-V4 Subjects: 1129 samples of heartbeats; 582 MI, 547 Normal | 85.71 |
| 10 | Safdarian et al., [41] | NB | NC | Leads: 12 leads Subjects: 52 Normal 148 MI from PTB database | 94.74 |
| 11 | Kora et al., [42] | IBA+LMNN | NC | Leads: Lead III Subjects: 52 Normal 148 MI from PTB database | 98.9 |
| 12 | Sharma et al., [43] | SVM-RBF | 10-fold CV | Leads: 12 leads Subject: 148 MI, 52 Normal from PTB database | 96.0 |
| 13 | Acharya et al., [63] | DWT Coefficients+KNN | 10-fold CV | Leads: 12 leads Subject: 52 normal, 148 MI from PTB database | 98.74 |
| 14 | Acharya et al., [64] | DCT Coefficients+KNN | 10-fold CV | Leads: Lead II Subject: 148 MI, 52 Normal from PTB database | 98.5 |
| 15 | Kumar et al., [85] | LS-SVM | 10-fold CV | Leads: Lead II subjects: 52 Normal and 148 MI from PTB database | 99.31 |
| 16 | Khatun and Morshed, [52] | BTs | 10-fold CV | Leads: 12 leads subjects: 79 normal, 346 MI from PTB database | 99.7 |
| 17 | Acharya et al., [65] | CWT-based controlet+KNN | 10-fold CV | Leads: 12 leads Subjects: 148 MI, 52 Normal from PTB database | 99.55 |
| 18 | Dohare et al., [66] | SVM+PCA | 10-fold CV | Leads: 12 leads subjects: 60 MI, 60 Normal from PTB database | 96.66 |
| 19 | Diker et al., [53] | GA+SVM | 10-fold CV | Leads: 12 leads subjects: 148 MI, 52 Normal from PTB database | 87.8 |
| 20 | Han and Shi, [67] | SVM-RBF | 10-fold CV | Leads: 12 leads subjects: 148 MI, 52 Normal from PTB database | 99.81 |
| 21 | Zhang et al. [54] | SAE+TB | 10-fold CV | Leads: Lead II subjects: 148 MI, 52 Normal from PTB database | 99.90 |
| 22 | Zeng et al., [68] | RBF- neural network | 10-fold CV | Leads: 12 leads Subjects: 148 MI, 52 Normal from PTB database | 97.98 |
| 23 | Kayikcioglu et al., [69] | Weighted KNN | 10-fold CV | leads: Leads V1-V5 Subjects: European ST-T (70 ECG recordings), MIT-BIH Arrhythmia database (46 different patients) and Long-Term ST (70 ECG recordings) | 94.23 |
| 24 | Liu et al. [70] | Dual-Q TQWT + DWPT + MPCA + TB | 10-fold CV | Leads: 12 leads and 3 Frank leads (VX, VY, VZ) Subjects: 78 Normal, 328 MI from PTB database | 97.46 |
| 25 | Lin et al., [44] | KNN | 10-fold CV | Leads: 12 Leads subjects: 148 MI and 52 Normal from PTB database | 99.57 |
| 26 | Valizadeh et al., [45] | Parametric-based technique + spherical harmonic-based feature selection +PCA+ SVM/KNN/RF | leave-one-out cross-validation scheme | 40 cine-MRI Subjects: 20 Normal and 20 MI from the automated cardiac diagnosis challenge database | 97.5 |
| 28 | Shahnawaz and Dawood, [86] | ANN | 10-fold CV | Leads: single lead Subjects: 148 MI, 52 Normal from PTB database | 99.1 |
| 29 | Panchavati et al., [87] | Gradient boosted tree model | 3-fold CV | Subjects: 253 MI, 1600 Normal, Electronic health records (age, glucose, systolic/diastolic blood pressure, etc.) | NC |
| 30 | Sulthana and Jaithunbi, [46] | Probabilistic PCA + multi-linear regression + RBF based SVMs | 5-fold CV | 517 tuples with 20 attributes, Clinical tests | 94.03 |
| 31 | Mohammad et al., [72] | ANN | NC | Subjects: 139288 MI in the SWEDEHEART registry and 30971 MI in the Western Denmark Heart Registry, Clinical data (age, sex, medical history, etc.) | NC |

**NC**: not considered; Other: rest of the publishers except IEEE, Springer, and Elsevier

**TABLE 2**
Summary of deep learning-based publications on myocardial infarct diagnosis

| No. | References | Methods | No. K-fold CV | Dataset | ACC (%) |
|---|---|---|---|---|---|
| 1 | Acharya et al., [73] | DCNN | 10-fold CV | Leads: Lead II subjects: 148 MI, 52 Normal from PTB database | 95.22 |

| No. | References | Methods | No. K-fold CV | Dataset | ACC (%) |
|---|---|---|---|---|---|
| 2 | Reasat and Shahnaz, [55] | DCNN | NC | Leads: Leads II, III and AVF<br>Subjects: 148 MI, 52 Normal from PTB database | 84.54 |
| 3 | Lui and Chow, [76] | DCNN-RNN | 10-fold CV | Leads: Lead I<br>Subjects: 148 MI and 52 Normal from PTB database | NC |
| 5 | Baloglu et al., [77] | DCNN | NC | Leads: 12 leads<br>Subjects: 52 Normal, 148 MI from PTB database | 99.78 |
| 6 | Tripathy et al., [56] | DL-LSSVM | 5-fold CV | Leads: 12 leads<br>Subjects: 148 MI, 52 Normal from PTB database | 99.74 |
| 7 | Zhang et al., [57] | GADF+PCANet+Linear SVM | 5-fold CV | Leads: Lead II<br>Subjects: 52 Normal, 148 MI from PTB database | 93.17 |
| 8 | Feng et al., [88] | 16-layer DCNN+LSTM | 10-fold CV | Leads: Lead I<br>Subjects: 148 MI, 52 Normal from PTB database | 95.40 |
| 9 | Liu et al., [58] | MFB-CBRNN | 5-fold CV | Leads: 12 leads<br>Subjects: 148 MI, 52 Normal from PTB database | 93.08 |
| 10 | Strodthoff and Strodthoff, [89] | DCNN-FC | 10-fold CV | Leads:12 leads<br>Subjects: 127 MI, 52 Normal | NC |
| 4 | Gupta et al., [47] | DCNNQuak | 100-fold CV | Leads: 12 leads along with 3 Frank leads<br>Subjects: 52 Normal, 148 MI from PTB database | 97.83 |
| 11 | Han and Shi, [74] | ML-ResNet | 5-fold CV | Leads: 12 leads<br>Subjects: 52 Normal, 113 MI from PTB database | 95.49 |
| 12 | Kim et al., [78] | AU-Net | NC | Leads: NC<br>Subjects: 14 coronary artery disease, 8 hypertrophic cardiomyopathy, and 13 Normal | NC |
| 13 | Natesan et al., [59] | DCNN+DA | NC | Leads: 12 leads<br>Subjects: 148 MI, 52 Normal from PTB database | 94.98 |
| 14 | Fu et al., [90] | MLA-DCNN-BiGRU | 5-fold CV | Leads: 12 leads<br>Subjects: 148 MI, 52 Normal from PTB database | 96.50 |
| 15 | Tadesse et al., [75] | Longitudinal+MnansNet | 10-fold CV | Leads: 12 leads<br>Subjects: 148 MI, 52 Normal from PTB database; 11853 MI, 5528 Normal from GCI database | 73.20 |
| 16 | Jahmunah et al., [32] | DCNN | 10-fold CV | Leads: Lead II<br>Subjects: 148 MI, 52 Normal from PTB database | 99.95 |
| 17 | Garland et al., [91] | InceptionResNet v2 | 5-fold CV | 150 images: 50 Normal,<br>50 Acute MI, and 50 Old MI,<br>Histology slides | 98.33 |

| No. | References | Methods | No. K-fold CV | Dataset | ACC (%) |
|---|---|---|---|---|---|
| 18 | Tadesse et al., [48] | GoogLeNet (Inception -v3)+Spectral-longitudinal model | 5-fold CV | Leads: 12 leads Subjects: 148 MI, 52 Normal from PTB database; 11853 MI, 5528 Normal from GCI database | NC |
| 19 | Jian et al., [92] | DCNNs | 5-fold CV | Leads: 12 leads Subjects: 148 MI, 52 Normal from PTB database; 11853 MI, 5528 Normal from GCI database | 95.76 |
| 20 | Wang et al., [93] | Multitask attention learning model | NC | 2414 hand images from 301 patients (182 males and 119 females), Hand images | 84.15 |
| 21 | Rai and Chatterjee, [49] | DCNN-LSTM + ensemble technique | NC | Leads: Lead IIs: 52 Normal, 148 MI from PTB database; 47 subjects from MIT-BIH arrhythmia database | 99.89 |
| 22 | Hammad et al., [50] | DCNN method based on focal loss | 5- FCV | Leads: 12 leads Subjects: 147 MI, 53 Normal from PTB database | 98.84 |
| 23 | He et al., [84] | DCNN + active learning | 5-fold CV | Leads: 12 leads Subjects: 113 MI, 52 Normal from PTB database, Subjects: 2784 MI, 1967 Normal from PTB-XL database | 96.99 |
| 24 | Yadav et al., [60] | DCNN | NC | Leads: 3 Frank leads, 148 records from PTB database | 99.82 |
| 25 | Degerli et al., [61] | Deep encoder-decoder CNN + SVM | 5-fold CV | 2349 images from 72 MI patients and 37 non-MI subjects, Echocardiography | 80.24 |
| 26 | Rai et al., [51] | DCNN+LSTM | NC | Leads: Lead II Subjects: 52 Normal, 148 MI from PTB database | 99.8 |
| 27 | Chen et al., [94] | DCNN-based residual network | NC | Leads: 12 leads 18,885 patients from PTB-XL database; 10,646 patients from Chapman University and Shaoxing People's Hospital | NC |
| 28 | Guo et al., [79] | Polar residual network | NC | Balanced dataset: 2,370 infarct, 2,370 normal; Unbalanced dataset: 2,370 infarct, 13,056 normal, Myocardial contrast echocardiography (MCE) | 98.7 |
| 29 | Xiong et al., [80] | Densely connected DCNN-based multi-lead localization method | 10-fold CV | Leads: 12 leads Subjects: 52 Normal, 148 MI from PTB database | 99.87 |
| 30 | Cao et al., [62] | DCNN-based multichannel lightweight model | 10-fold CV | Leads: Leads V2, V3, V5, aVL Subjects: 147 MI, 53 Normal from PTB database | 96.65 |
| 31 | Borisov et al., [95] | Linear SVM+ PCA | NC | Leads: 12 leads 30 MI, 42 normal | NC |
| 32 | Chen et al., [81] | Framework A (Baseline CNN) + Framework B (Context Encoder Network) | NC | 904 Delayed Enhancement cardiac MRI slices | 66.8 |

| No. | References | Methods | No. K-fold CV | Dataset | ACC (%) |
|---|---|---|---|---|---|
| 33 | Liu et al., [96] | Evolving Multi-branch Network | 5- FCV | Leads: 12-leads Subjects: 148 MI, 52 Normal from PTB database; 21,837 records from the PTB-XL database | 97.11 |
| 34 | Wang et al., [82] | integrating deep auto-weighted supervision and pixel-wise attention network | NC | Multi-sequence cardiac MRI, including bSSFP, LGE, T2, from 45 patients | 70.7 |
| 35 | Li et al., [83] | GAN + DCNN | 5-fold CV | Leads: single lead ECG from PTB Diagnostic ECG Database | 99.06 |

**NC**: not considered

More ML works were published previously, but DL publications are gradually superseding the numbers in recent years. In 2021, there were 17 DL vs. 4 ML publications on MI diagnosis. The secular trend of ML and DL publications is shown in Figs. 5 and 6, respectively. Hence, the number of papers on MI diagnosis using DL-based methods has increased in recent years. Even though DL-based MI detection began later than ML-based detection, the number of DL publications has caught up with ML publications; 35 papers for DL and 31 papers for ML (Fig. 7a). However, the model performance of ML publications is more consistent than that of DL publications. The box-and-whiskers plot in Fig.7b shows that the model performance of ML-based MI detection has a lower standard deviation and the range of accuracy scores falls between 79.0 to 99.9%, while the range of accuracy score for DL-based MI detection is 66.8 to 99.95%.

Despite many studies proposing various ML/DL approaches for medical applications; ML/DL still suffers from some limitations. First of all, medical datasets may contain samples with missing values. These samples are not readily usable during ML/DL model training. Avoiding samples with missing values causes biased [102] training/evaluation of models, which is not desirable. DL methods have huge potential for knowledge learning and representation, but only if a sufficient number of training samples are fed. In the medical domain, gathering and labeling a large number of samples is usually challenging [103]. On the other hand, a limited number of training samples causes DL models to underperform in the test phase. Furthermore, DL models are not error-free, and wrong predictions can be catastrophic in medical applications. Therefore, ML/DL models must be able to determine whether their outputs are trustworthy or not. Unfortunately, not all models are equipped with such ability. ML/DL community has come a long way. However, current ML/DL methods are still not robust enough to fully gain medical expert's trust. Therefore, ML/DL application in medical domain is still limited.

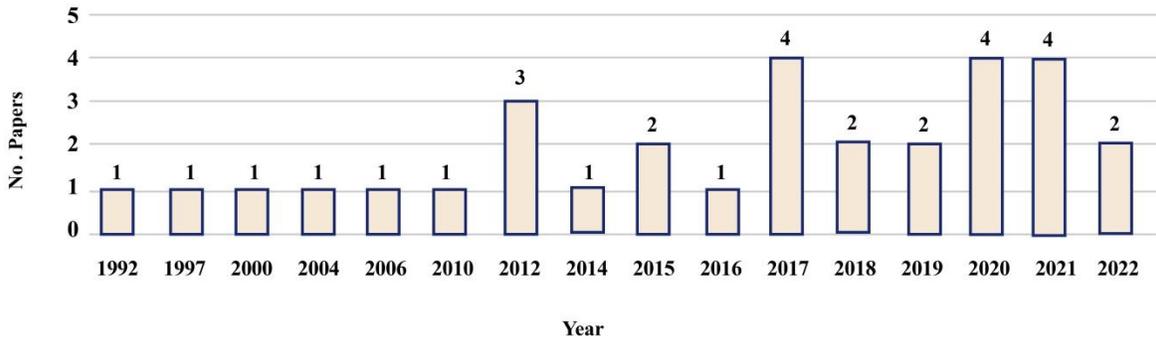

Fig. 5. The number of published papers on myocardial infarct diagnosis using machine learning-based methods between 1992 and 2022.

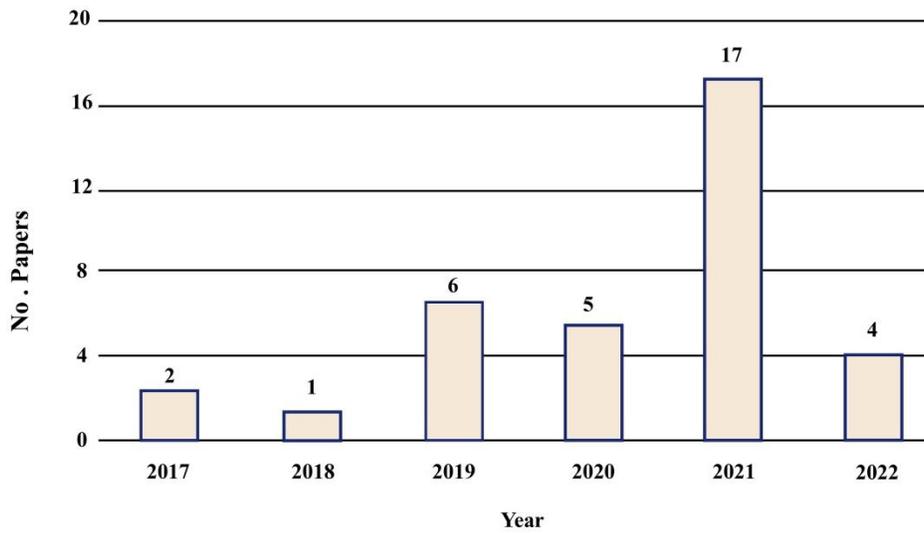

Fig. 6. Number of papers for myocardial infarct diagnosis using deep learning-based methods between 2017 and 2022.

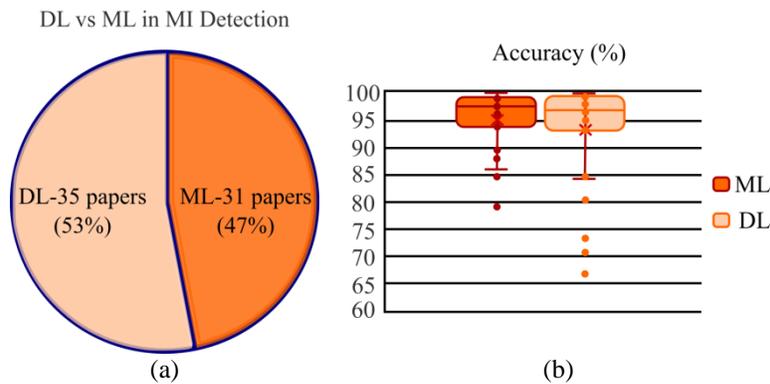

Fig. 7. (a) Pie chart representation on the number of DL publications versus ML publications in MI detection (b) Box-and-whiskers plot of model accuracy score for DL and ML publications.

## V. FUTURE WORK

In recent years, the outstanding representation power of DL has led to the rapid growth of DL-based studies for MI detection. However, DL-based approaches demand high computational power and massive memory that may not be available in all medical centers. Therefore, striving to make DL methods cloud-compatible is a stepping stone toward the wide application of DL in a clinical setting. Currently, the limitation of DL is its massive memory consumption, which makes cloud storage impractical [104]. Hence, as future work on DL-based MI detection, it is desirable to develop practical clinical decision support tools capable of being used both in and out of the hospital, like in Fig 8. As can be seen, in the setup of Fig 8, wearable devices act as an interface between the patient and remote medical services.

Additionally, further improvement and reliability of MI detection using DL is also a possible direction for future works. Furthermore, it is desirable to reduce the time of input signals preparation and preprocessing. Heart rate signals extracted from ECG can be used for MI detection [4, 105]. The heart rate signals demand lower bandwidth, so using those yields a significant reduction in memory requirement. Alternatively, heart rate signals may be obtained from photoplethysmography signals [105] acquired using wearable devices (e.g. wristwatch).

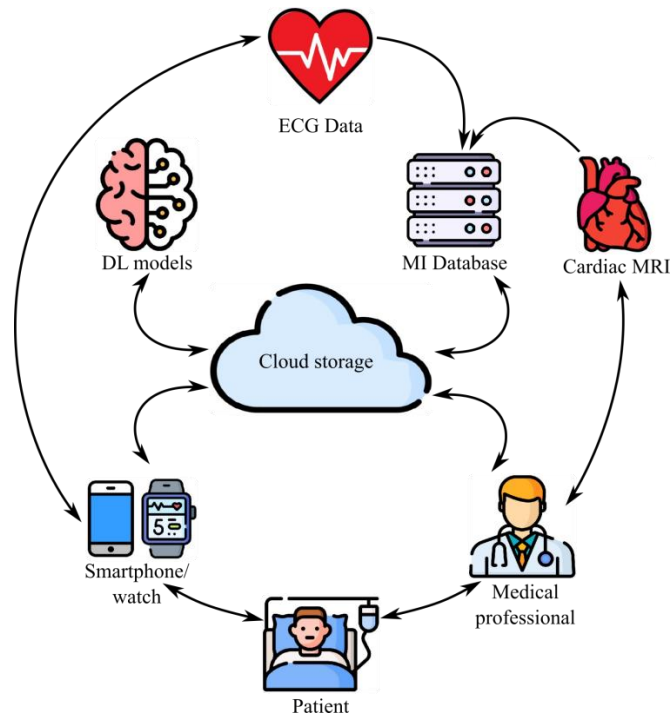

**Fig. 8.** Ideal cloud-based system for MI detection.

## VI. Conclusion and closing thoughts

ECG signals are frequently used to screen for MI. On the other hand, manual ECG is time-consuming and subject to bias. To overcome the aforementioned challenges in MI diagnosis, ML and DL methods can be used. ML methods rely on handcrafted features based on ECG signals, whereas DL is capable of automatic feature extraction. We reviewed the methods based on ML and DL for MI diagnosis. To this end, several papers were collected based on search keywords. Thirty-one papers focused on ML methods and thirty-five on DL methods. According to the reviewed papers, DCNN models yield the highest accuracy for MI diagnosis in DL. As a result, many researchers have used DL methods in recent years. Nevertheless, as with any other method DL has its drawbacks as well. A high number of training samples and heavy computation load during training are two major drawbacks of DL methods. There is ongoing research on the reduction of the computational complexity of DL models and making them more data-efficient. For example, knowledge distillation approaches can be used to run DL models on embedded devices with limited computation power. The knowledge distillation process has two main phases. First, a complex DL model is trained and tuned. In the second phase, a lightweight DL model is trained to mimic the behavior of the complex model while accepting performance degradation to a reasonable extent. Regarding data efficiency, various data augmentation methods can be used. Moreover, generative models such as generative adversarial networks are also a viable solution.

## References

[1] L. Sun, Y. Lu, K. Yang, S. Li, ECG analysis using multiple instance learning for myocardial infarction detection, IEEE transactions on biomedical engineering, 59 (2012) 3348-3356.

[2] J.G. Canto, M.G. Shlipak, W.J. Rogers, J.A. Malmgren, P.D. Frederick, C.T. Lambrew, J.P. Ornato, H.V. Barron, C.I. Kiefe, Prevalence, clinical characteristics, and mortality among patients with myocardial infarction presenting without chest pain, Jama, 283 (2000) 3223-3229.

[3] E. Boersma, N. Mercado, D. Poldermans, M. Gardien, J. Vos, M.L. Simoons, Acute myocardial infarction, The Lancet, 361 (2003) 847-858.

[4] E. Jayachandran, Analysis of myocardial infarction using discrete wavelet transform, Journal of medical systems, 34 (2010) 985-992.

[5] H. Yang, C. Kan, G. Liu, Y. Chen, Spatiotemporal differentiation of myocardial infarctions, IEEE Transactions on Automation Science and Engineering, 10 (2013) 938-947.

[6] E.E. Creemers, J.P. Cleutjens, J.F. Smits, M.J. Daemen, Matrix metalloproteinase inhibition after myocardial infarction: a new approach to prevent heart failure?, Circulation research, 89 (2001) 201-210.

[7] B. Liu, J. Liu, G. Wang, K. Huang, F. Li, Y. Zheng, Y. Luo, F. Zhou, A novel electrocardiogram parameterization algorithm and its application in myocardial infarction detection, Computers in biology and medicine, 61 (2015) 178-184.

[8] M. Hammad, K.N. Rajesh, A. Abdelatey, M. Abdar, M. Zomorodi-Moghadam, R. San Tan, U.R. Acharya, J. Pławiak, R. Tadeusiewicz, V. Makarenkov, Automated detection of Shockable ECG signals: A Review, Information Sciences, (2021).

[9] S. Banerjee, M. Mitra, Cross wavelet transform based analysis of electrocardiogram signals, International Journal of Electrical, Electronics and Computer Engineering, 1 (2012) 88-92.

[10] F. Jiang, Y. Jiang, H. Zhi, Y. Dong, H. Li, S. Ma, Y. Wang, Q. Dong, H. Shen, Y. Wang, Artificial intelligence in healthcare: past, present and future, Stroke and vascular neurology, 2 (2017).

[11] K.-H. Yu, A.L. Beam, I.S. Kohane, Artificial intelligence in healthcare, Nature biomedical engineering, 2 (2018) 719-731.

[12] T. Davenport, R. Kalakota, The potential for artificial intelligence in healthcare, Future healthcare journal, 6 (2019) 94.

[13] J.M. Ribeiro, P. Astudillo, O. de Backer, R. Budde, R.J. Nuis, J. Goudzwaard, N.M. Van Mieghem, J. Lumens, P. Mortier, F.M. Raso, Artificial Intelligence and Transcatheter Interventions for Structural Heart Disease: A glance at the (near) future, Trends in Cardiovascular Medicine, (2021).
[14] T.M. Mitchell, Machine learning, (1997).
[15] A.S. Alharthi, S.U. Yunas, K.B. Ozanyan, Deep learning for monitoring of human gait: A review, IEEE Sensors Journal, 19 (2019) 9575-9591.
[16] O.S. Lih, V. Jahmunah, T.R. San, E.J. Ciaccio, T. Yamakawa, M. Tanabe, M. Kobayashi, O. Faust, U.R. Acharya, Comprehensive electrocardiographic diagnosis based on deep learning, Artificial intelligence in medicine, 103 (2020) 101789.
[17] A.S. Lundervold, A. Lundervold, An overview of deep learning in medical imaging focusing on MRI, Zeitschrift für Medizinische Physik, 29 (2019) 102-127.
[18] J.H. Joloudari, M. Haderbadi, A. Mashmool, M. GhasemiGol, S.S. Band, A. Mosavi, Early detection of the advanced persistent threat attack using performance analysis of deep learning, IEEE Access, 8 (2020) 186125-186137.
[19] S. Nosratabadi, A. Mosavi, P. Duan, P. Ghamisi, F. Filip, S.S. Band, U. Reuter, J. Gama, A.H. Gandomi, Data science in economics: comprehensive review of advanced machine learning and deep learning methods, Mathematics, 8 (2020) 1799.
[20] T.-H. Huang, W.-K. Lee, C.-C. Wu, C.-C. Lee, C.-F. Lu, H.-C. Yang, C.-Y. Lin, W.-Y. Chung, P.-S. Wang, Y.-L. Chen, Detection of Vestibular Schwannoma on Triple-parametric Magnetic Resonance Images Using Convolutional Neural Networks, Journal of Medical and Biological Engineering, (2021) 1-10.
[21] F. Shi, J. Wang, J. Shi, Z. Wu, Q. Wang, Z. Tang, K. He, Y. Shi, D. Shen, Review of artificial intelligence techniques in imaging data acquisition, segmentation, and diagnosis for COVID-19, IEEE reviews in biomedical engineering, 14 (2020) 4-15.
[22] A.S. Panayides, A. Amini, N.D. Filipovic, A. Sharma, S.A. Tsaftaris, A. Young, D. Foran, N. Do, S. Golemati, T. Kurc, AI in medical imaging informatics: current challenges and future directions, IEEE Journal of Biomedical and Health Informatics, 24 (2020) 1837-1857.
[23] O. Faust, Y. Hagiwara, T.J. Hong, O.S. Lih, U.R. Acharya, Deep learning for healthcare applications based on physiological signals: A review, Computer methods and programs in biomedicine, 161 (2018) 1-13.
[24] D. Shen, G. Wu, H.-I. Suk, Deep learning in medical image analysis, Annual review of biomedical engineering, 19 (2017) 221-248.
[25] J. Ker, L. Wang, J. Rao, T. Lim, Deep learning applications in medical image analysis, Ieee Access, 6 (2017) 9375-9389.
[26] G. Litjens, T. Kooi, B.E. Bejnordi, A.A.A. Setio, F. Ciompi, M. Ghafoorian, J.A. Van Der Laak, B. Van Ginneken, C.I. Sánchez, A survey on deep learning in medical image analysis, Medical image analysis, 42 (2017) 60-88.
[27] M. Chen, X. Shi, Y. Zhang, D. Wu, M. Guizani, Deep features learning for medical image analysis with convolutional autoencoder neural network, IEEE Transactions on Big Data, (2017).
[28] W. Rawat, Z. Wang, Deep convolutional neural networks for image classification: A comprehensive review, Neural computation, 29 (2017) 2352-2449.
[29] A. Esteva, A. Robicquet, B. Ramsundar, V. Kuleshov, M. DePristo, K. Chou, C. Cui, G. Corrado, S. Thrun, J. Dean, A guide to deep learning in healthcare, Nature medicine, 25 (2019) 24-29.
[30] F. Ali, S. El-Sappagh, S.R. Islam, D. Kwak, A. Ali, M. Imran, K.-S. Kwak, A smart healthcare monitoring system for heart disease prediction based on ensemble deep learning and feature fusion, Information Fusion, 63 (2020) 208-222.
[31] Y. Sun, B. Xue, M. Zhang, G.G. Yen, J. Lv, Automatically designing CNN architectures using the genetic algorithm for image classification, IEEE transactions on cybernetics, 50 (2020) 3840-3854.
[32] V. Jahmunah, E. Ng, T.R. San, U.R. Acharya, Automated detection of coronary artery disease, myocardial infarction and congestive heart failure using GaborCNN model with ECG signals, Computers in biology and medicine, 134 (2021) 104457.
[33] M. Reddy, L. Edenbrandt, J. Svensson, W. Haisty, O. Pahlm, Neural network versus electrocardiographer and conventional computer criteria in diagnosing anterior infarct from the ECG, Proceedings Computers in Cardiology, IEEE, 1992, pp. 667-670.
[34] B. Hedén, H. Ohlin, R. Rittner, L. Edenbrandt, Acute myocardial infarction detected in the 12-lead ECG by artificial neural networks, Circulation, 96 (1997) 1798-1802.
[35] H. Lu, K. Ong, P. Chia, An automated ECG classification system based on a neuro-fuzzy system, Computers in Cardiology 2000. Vol. 27 (Cat. 00CH37163), IEEE, 2000, pp. 387-390.
[36] H. Haraldsson, L. Edenbrandt, M. Ohlsson, Detecting acute myocardial infarction in the 12-lead ECG using Hermite expansions and neural networks, Artificial Intelligence in Medicine, 32 (2004) 127-136.
[37] H. Zheng, H. Wang, C. Nugent, D. Finlay, Supervised classification models to detect the presence of old myocardial infarction in body surface potential maps, 2006 Computers in Cardiology, IEEE, 2006, pp. 265-268.
[38] M. Arif, I.A. Malagore, F.A. Afsar, Automatic detection and localization of myocardial infarction using back propagation neural networks, 2010 4th International Conference on Bioinformatics and Biomedical Engineering, IEEE, 2010, pp. 1-4.
[39] M. Arif, I.A. Malagore, F.A. Afsar, Detection and localization of myocardial infarction using k-nearest neighbor classifier, Journal of medical systems, 36 (2012) 279-289.
[40] P.-C. Chang, J.-J. Lin, J.-C. Hsieh, J. Weng, Myocardial infarction classification with multi-lead ECG using hidden Markov models and Gaussian mixture models, Applied Soft Computing, 12 (2012) 3165-3175.
[41] N. Safdarian, N.J. Dabanloo, G. Attarodi, A new pattern recognition method for detection and localization of myocardial infarction using T-wave integral and total integral as extracted features from one cycle of ECG signal, Journal of Biomedical Science and Engineering, 2014 (2014).
[42] P. Kora, S.R. Kalva, Improved Bat algorithm for the detection of myocardial infarction, SpringerPlus, 4 (2015) 666.
[43] L. Sharma, R. Tripathy, S. Dandapat, Multiscale energy and eigenspace approach to detection and localization of myocardial infarction, IEEE transactions on biomedical engineering, 62 (2015) 1827-1837.
[44] Z. Lin, Y. Gao, Y. Chen, Q. Ge, G. Mahara, J. Zhang, Automated detection of myocardial infarction using robust features extracted from 12-lead ECG, Signal, Image and Video Processing, (2020) 1-9.
[45] G. Valizadeh, F.B. Mofrad, A. Shalbaf, Parametric-based feature selection via spherical harmonic coefficients for the left ventricle myocardial infarction screening, Medical & Biological Engineering & Computing, (2021) 1-23.
[46] A.R. Sulthana, A. Jaithunbi, Varying combination of feature extraction and modified support vector machines based prediction of myocardial infarction, Evolving Systems, (2022) 1-18.
[47] A. Gupta, E. Huerta, Z. Zhao, I. Moussa, Deep Learning for Cardiologist-level Myocardial Infarction Detection in Electrocardiograms, European Medical and Biological Engineering Conference, Springer, 2020, pp. 341-355.
[48] G.A. Tadesse, K. Weldemariam, H. Javed, Y. Liu, J. Liu, J. Chen, T. Zhu, Discriminant Knowledge Extraction from Electrocardiograms for Automated Diagnosis of Myocardial Infarction, Knowledge Management and Acquisition for Intelligent Systems: 17th Pacific Rim Knowledge Acquisition Workshop, PKAW 2020, Yokohama, Japan, January 7–8, 2021, Proceedings 17, Springer International Publishing, 2021, pp. 70-82.
[49] H.M. Rai, K. Chatterjee, Hybrid CNN-LSTM deep learning model and ensemble technique for automatic detection of myocardial infarction using big ECG data, Applied Intelligence, (2021) 1-19.
[50] M. Hammad, M.H. Alkinani, B. Gupta, A.A. Abd El-Latif, Myocardial infarction detection based on deep neural network on imbalanced data, Multimedia Systems, (2021) 1-13.
[51] H.M. Rai, K. Chatterjee, A. Dubey, P. Srivastava, Myocardial Infarction Detection Using Deep Learning and Ensemble Technique from ECG Signals, Proceedings of Second International Conference on Computing, Communications, and Cyber-Security, Springer, 2021, pp. 717-730.
[52] S. Khatun, B.I. Morshed, Detection of myocardial infarction and arrhythmia from single-lead ECG data using bagging trees classifier, 2017 IEEE International Conference on Electro Information Technology (EIT), IEEE, 2017, pp. 520-524.

[53] A. Diker, Z. Cömert, E. Avci, S. Velappan, Intelligent system based on Genetic Algorithm and support vector machine for detection of myocardial infarction from ECG signals,  2018 26th Signal Processing and Communications Applications Conference (SIU), IEEE, 2018, pp. 1-4.
[54] J. Zhang, F. Lin, P. Xiong, H. Du, H. Zhang, M. Liu, Z. Hou, X. Liu, Automated detection and localization of myocardial infarction with staked sparse autoencoder and treebagger, IEEE Access, 7 (2019) 70634-70642.
[55] T. Reasat, C. Shahnaz, Detection of inferior myocardial infarction using shallow convolutional neural networks,  2017 IEEE Region 10 Humanitarian Technology Conference (R10-HTC), IEEE, 2017, pp. 718-721.
[56] R.K. Tripathy, A. Bhattacharyya, R.B. Pachori, A novel approach for detection of myocardial infarction from ECG signals of multiple electrodes, IEEE Sensors Journal, 19 (2019) 4509-4517.
[57] G. Zhang, Y. Si, D. Wang, W. Yang, Y. Sun, Automated detection of myocardial infarction using a gramian angular field and principal component analysis network, IEEE Access, 7 (2019) 171570-171583.
[58] W. Liu, F. Wang, Q. Huang, S. Chang, H. Wang, J. He, MFB-CBRNN: A hybrid network for MI detection using 12-lead ECGs, IEEE journal of biomedical and health informatics, 24 (2019) 503-514.
[59] P. Natesan, E. Gothai, Classification of Multi-Lead ECG Signals to Predict Myocardial Infarction Using CNN,  2020 Fourth International Conference on Computing Methodologies and Communication (ICCMC), IEEE, 2020, pp. 1029-1033.
[60] S.S. Yadav, S.B. More, S.M. Jadhav, S.R. Sutar, Convolutional neural networks based diagnosis of myocardial infarction in electrocardiograms,  2021 International Conference on Computing, Communication, and Intelligent Systems (ICCCIS), IEEE, 2021, pp. 581-586.
[61] A. Degerli, M. Zabihi, S. Kiranyaz, T. Hamid, R. Mazhar, R. Hamila, M. Gabbouj, Early Detection of Myocardial Infarction in Low-Quality Echocardiography, IEEE Access, 9 (2021) 34442-34453.
[62] Y. Cao, T. Wei, B. Zhang, N. Lin, J.J. Rodrigues, J. Li, D. Zhang, ML-Net: Multi-Channel Lightweight Network for Detecting Myocardial Infarction, IEEE Journal of Biomedical and Health Informatics, (2021).
[63] U.R. Acharya, H. Fujita, V.K. Sudarshan, S.L. Oh, M. Adam, J.E. Koh, J.H. Tan, D.N. Ghista, R.J. Martis, C.K. Chua, Automated detection and localization of myocardial infarction using electrocardiogram: a comparative study of different leads, Knowledge-Based Systems, 99 (2016) 146-156.
[64] U.R. Acharya, H. Fujita, M. Adam, O.S. Lih, V.K. Sudarshan, T.J. Hong, J.E. Koh, Y. Hagiwara, C.K. Chua, C.K. Poo, Automated characterization and classification of coronary artery disease and myocardial infarction by decomposition of ECG signals: A comparative study, Information Sciences, 377 (2017) 17-29.
[65] U.R. Acharya, H. Fujita, V.K. Sudarshan, S.L. Oh, M. Adam, J.H. Tan, J.H. Koo, A. Jain, C.M. Lim, K.C. Chua, Automated characterization of coronary artery disease, myocardial infarction, and congestive heart failure using contourlet and shearlet transforms of electrocardiogram signal, Knowledge-Based Systems, 132 (2017) 156-166.
[66] A.K. Dohare, V. Kumar, R. Kumar, Detection of myocardial infarction in 12 lead ECG using support vector machine, Applied Soft Computing, 64 (2018) 138-147.
[67] C. Han, L. Shi, Automated interpretable detection of myocardial infarction fusing energy entropy and morphological features, Computer methods and programs in biomedicine, 175 (2019) 9-23.
[68] W. Zeng, J. Yuan, C. Yuan, Q. Wang, F. Liu, Y. Wang, Classification of myocardial infarction based on hybrid feature extraction and artificial intelligence tools by adopting tunable-Q wavelet transform (TQWT), variational mode decomposition (VMD) and neural networks, Artificial Intelligence in Medicine, (2020) 101848.
[69] İ. Kayikcioglu, F. Akdeniz, C. Köse, T. Kayikcioglu, Time-frequency approach to ECG classification of myocardial infarction, Computers & Electrical Engineering, 84 (2020) 106621.
[70] J. Liu, C. Zhang, Y. Zhu, T. Ristaniemi, T. Parviainen, F. Cong, Automated detection and localization system of myocardial infarction in single-beat ECG using Dual-Q TQWT and wavelet packet tensor decomposition, Computer methods and programs in biomedicine, 184 (2020) 105120.
[71] E. Avard, I. Shiri, G. Hajianfar, H. Abdollahi, K.R. Kalantari, G. Houshmand, K. Kasani, A. Bitarafan-Rajabi, M.R. Deevband, M. Oveisi, Non-contrast Cine Cardiac Magnetic Resonance image radiomics features and machine learning algorithms for myocardial infarction detection, Computers in biology and medicine, (2021) 105145.
[72] M.A. Mohammad, K.K. Olesen, S. Koul, C.P. Gale, R. Rylance, T. Jernberg, T. Baron, J. Spaak, S. James, B. Lindahl, Development and validation of an artificial neural network algorithm to predict mortality and admission to hospital for heart failure after myocardial infarction: a nationwide population-based study, The Lancet Digital Health, 4 (2022) e37-e45.
[73] U.R. Acharya, H. Fujita, S.L. Oh, Y. Hagiwara, J.H. Tan, M. Adam, Application of deep convolutional neural network for automated detection of myocardial infarction using ECG signals, Information Sciences, 415 (2017) 190-198.
[74] C. Han, L. Shi, ML–ResNet: A novel network to detect and locate myocardial infarction using 12 leads ECG, Computer methods and programs in biomedicine, 185 (2020) 105138.
[75] G.A. Tadesse, H. Javed, K. Weldemariam, Y. Liu, J. Liu, J. Chen, T. Zhu, DeepMI: Deep multi-lead ECG fusion for identifying myocardial infarction and its occurrence-time, Artificial Intelligence in Medicine, 121 (2021) 102192.
[76] H.W. Lui, K.L. Chow, Multiclass classification of myocardial infarction with convolutional and recurrent neural networks for portable ECG devices, Informatics in Medicine Unlocked, 13 (2018) 26-33.
[77] U.B. Baloglu, M. Talo, O. Yildirim, R. San Tan, U.R. Acharya, Classification of myocardial infarction with multi-lead ECG signals and deep CNN, Pattern Recognition Letters, 122 (2019) 23-30.
[78] Y.-C. Kim, K.R. Kim, Y.H. Choe, Automatic myocardial segmentation in dynamic contrast enhanced perfusion MRI using Monte Carlo dropout in an encoder-decoder convolutional neural network, Computer methods and programs in biomedicine, 185 (2020) 105150.
[79] Y. Guo, G.-Q. Du, W.-Q. Shen, C. Du, P.-N. He, S. Siuly, Automatic myocardial infarction detection in contrast echocardiography based on polar residual network, Computer Methods and Programs in Biomedicine, 198 (2021) 105791.
[80] P. Xiong, Y. Xue, J. Zhang, M. Liu, H. Du, H. Zhang, Z. Hou, H. Wang, X. Liu, Localization of myocardial infarction with multi-lead ECG based on DenseNet, Computer Methods and Programs in Biomedicine, 203 (2021) 106024.
[81] Z. Chen, A. Lalande, M. Salomon, T. Decourselle, T. Pommier, A. Qayyum, J. Shi, G. Perrot, R. Couturier, Automatic deep learning-based myocardial infarction segmentation from delayed enhancement MRI, Computerized Medical Imaging and Graphics, 95 (2022) 102014.
[82] K.-N. Wang, X. Yang, J. Miao, L. Li, J. Yao, P. Zhou, W. Xue, G.-Q. Zhou, X. Zhuang, D. Ni, AWSnet: An Auto-weighted Supervision Attention Network for Myocardial Scar and Edema Segmentation in Multi-sequence Cardiac Magnetic Resonance Images, Medical Image Analysis, (2022) 102362.
[83] W. Li, Y.M. Tang, K.M. Yu, S. To, SLC-GAN: An Automated Myocardial Infarction Detection Model Based on Generative Adversarial Networks and Convolutional Neural Networks with Single-Lead Electrocardiogram Synthesis, Information Sciences, (2022).
[84] Z. He, Z. Yuan, P. An, J. Zhao, B. Du, MFB-LANN: A lightweight and updatable myocardial infarction diagnosis system based on convolutional neural networks and active learning, Computer Methods and Programs in Biomedicine, 210 (2021) 106379.
[85] M. Kumar, R.B. Pachori, U.R. Acharya, Automated diagnosis of myocardial infarction ECG signals using sample entropy in flexible analytic wavelet transform framework, Entropy, 19 (2017) 488.
[86] M.B. Shahnawaz, H. Dawood, An Effective Deep Learning Model for Automated Detection of Myocardial Infarction Based on Ultrashort-Term Heart Rate Variability Analysis, Mathematical Problems in Engineering, 2021 (2021).
[87] S. Panchavati, C. Lam, N.S. Zelin, E. Pellegrini, G. Barnes, J. Hoffman, A. Garikipati, J. Calvert, Q. Mao, R. Das, Retrospective validation of a machine learning clinical decision support tool for myocardial infarction risk stratification, Healthcare technology letters, 8 (2021) 139.

[88] K. Feng, X. Pi, H. Liu, K. Sun, Myocardial infarction classification based on convolutional neural network and recurrent neural network, Applied Sciences, 9 (2019) 1879.
[89] N. Strodthoff, C. Strodthoff, Detecting and interpreting myocardial infarction using fully convolutional neural networks, Physiological measurement, 40 (2019) 015001.
[90] L. Fu, B. Lu, B. Nie, Z. Peng, H. Liu, X. Pi, Hybrid network with attention mechanism for detection and location of myocardial infarction based on 12-lead electrocardiogram signals, Sensors, 20 (2020) 1020.
[91] J. Garland, M. Hu, M. Duffy, K. Kesha, C. Glenn, P. Morrow, S. Stables, B. Ondruschka, U. Da Broi, R.D. Tse, Classifying Microscopic Acute and Old Myocardial Infarction Using Convolutional Neural Networks, The American Journal of Forensic Medicine and Pathology, 42 (2021) 230-234.
[92] J.-Z. Jian, T.-R. Ger, H.-H. Lai, C.-M. Ku, C.-A. Chen, P.A.R. Abu, S.-L. Chen, Detection of Myocardial Infarction Using ECG and Multi-Scale Feature Concatenate, Sensors, 21 (2021) 1906.
[93] Q. Wang, C. Zhao, Y. Qiang, Z. Zhao, K. Song, S. Luo, Multitask Interactive Attention Learning Model Based on Hand Images for Assisting Chinese Medicine in Predicting Myocardial Infarction, Computational and Mathematical Methods in Medicine, 2021 (2021).
[94] X. Chen, W. Guo, L. Zhao, W. Huang, L. Wang, A. Sun, L. Li, F. Mo, Acute myocardial infarction detection using deep learning-enabled electrocardiograms, Frontiers in cardiovascular medicine, 8 (2021).
[95] A.V. Borisov, A.G. Syrkina, D.A. Kuzmin, V.V. Ryabov, A.A. Boyko, O. Zaharova, V.S. Zasedatel, Y.V. Kistenev, Application of machine learning and laser optical-acoustic spectroscopy to study the profile of exhaled air volatile markers of acute myocardial infarction, Journal of Breath Research, 15 (2021) 027104.
[96] W. Liu, J. Ji, S. Chang, H. Wang, J. He, Q. Huang, EvoMBN: Evolving Multi-Branch Networks on Myocardial Infarction Diagnosis Using 12-Lead Electrocardiograms, Biosensors, 12 (2022) 15.
[97] D. Sharifrazi, R. Alizadehsani, J.H. Joloudari, S. Shamshirband, S. Hussain, Z.A. Sani, F. Hasanzadeh, A. Shoaibi, A. Dehzangi, H. Alinejad-Rokny, CNN-KCL: Automatic Myocarditis Diagnosis using Convolutional Neural Network Combined with K-means Clustering, (2020).
[98] L.D. Sharma, R.K. Sunkaria, Inferior myocardial infarction detection using stationary wavelet transform and machine learning approach, Signal, Image and Video Processing, 12 (2018) 199-206.
[99] A.L. Goldberger, L.A. Amaral, L. Glass, J.M. Hausdorff, P.C. Ivanov, R.G. Mark, J.E. Mietus, G.B. Moody, C.-K. Peng, H.E. Stanley, PhysioBank, PhysioToolkit, and PhysioNet: components of a new research resource for complex physiologic signals, circulation, 101 (2000) e215-e220.
[100] P. Wagner, N. Strodthoff, R.-D. Bousseljot, D. Kreiseler, F.I. Lunze, W. Samek, T. Schaeffter, PTB-XL, a large publicly available electrocardiography dataset, Scientific data, 7 (2020) 1-15.
[101] G.B. Moody, R.G. Mark, The MIT-BIH arrhythmia database on CD-ROM and software for use with it,  [1990] Proceedings Computers in Cardiology, IEEE, 1990, pp. 185-188.
[102] J. Wallert, M. Tomasoni, G. Madison, C. Held, Predicting two-year survival versus non-survival after first myocardial infarction using machine learning and Swedish national register data, BMC medical informatics and decision making, 17 (2017) 1-11.
[103] M.P. Than, J.W. Pickering, Y. Sandoval, A.S. Shah, A. Tsanas, F.S. Apple, S. Blankenberg, L. Cullen, C. Mueller, J.T. Neumann, Machine learning to predict the likelihood of acute myocardial infarction, Circulation, 140 (2019) 899-909.
[104] I. Tobore, J. Li, L. Yuhang, Y. Al-Handarish, A. Kandwal, Z. Nie, L. Wang, Deep learning intervention for health care challenges: some biomedical domain considerations, JMIR mHealth and uHealth, 7 (2019) e11966.
[105] H.W. Loh, S. Xu, O. Faust, C.P. Ooi, P.D. Barua, S. Chakraborty, R.-S. Tan, F. Molinari, U.R. Acharya, Application of Photoplethysmography signals for Healthcare systems: An in-depth review, Computer Methods and Programs in Biomedicine, (2022) 106677.